\documentclass[12pt, a4paper]{article}

\usepackage[utf8]{inputenc}
\usepackage{amsfonts,amsmath,amssymb}
\usepackage{bbold}
\usepackage{multicol}
\usepackage{cite}
\usepackage{slashed}
\usepackage{tikz}
\usepackage[compat=1.1.0]{tikz-feynman}
\usepackage{bigints}

\usepackage[
   top=10mm,
   bottom=20mm,
   left=20mm,
   right=20mm]{geometry}

\usepackage{xcolor}
\usepackage{authblk,ulem}
\usepackage{hyperref} 
\usepackage{caption}
\usepackage{graphicx}

\newcommand{\beq}{\begin{equation}}
\newcommand{\eeq}{\end{equation}}
\newcommand{\ba}{\begin{array}}
\newcommand{\ea}{\end{array}}
\newcommand{\bea}{\begin{eqnarray}}
\newcommand{\eea}{\end{eqnarray} }
\newcommand{\bal}{\begin{align}}
\newcommand{\eal}{\end{align}}

\title{Infrared Subtleties and Chiral Vertices at NLO: An Implicit Regularization Analysis}

\author[1]{Ricardo J. C. Rosado}

\author[2,3]{Adriano Cherchiglia}

\author[4]{Marcos Sampaio}

\author[1]{Brigitte Hiller}

\affil[1]{CFisUC, Department of Physics, University of Coimbra, P-3004-516 Coimbra,
Portugal}
\affil[2]{Instituto de Física Gleb Wataghin, Universidade Estadual de Campinas, \\ Rua Sérgio Buarque de Holanda, 777, Campinas, SP, Brasil}
\affil[3]{Departamento de F\'isica Te\'orica y del Cosmos, Universidad de Granada, Campus de Fuentenueva, E–18071 Granada, Spain}
\affil[4]{ Universidade Federal do ABC, 09210-580 , Santo Andr\'e, Brasil}

\begin{document}

\maketitle

\begin{abstract}
    We employ implicit regularization (IReg) in quark-antiquark decays of the Z, or of a scalar (CP-even or odd) boson at NLO, and compare with dimensional schemes to reveal subtleties involving infrared divergence cancellation and $\gamma_5$-matrix issues. Besides the absence of evanescent fields in IReg, such as $\epsilon$-scalars required in certain schemes that operate partially in the physical dimension, we verify that our procedure preserves gauge invariance in the presence of the $\gamma_5$ matrix without requiring symmetry preserving counterterms while the amplitude is infrared finite as stated by the KLN theorem.  
\end{abstract}

\section{Introduction}

Different regularization frameworks have been used in quantum field theory. Each of these frameworks has its advantages and disadvantages, and the choice of the appropriate framework depends on the specific problem being considered \cite{gnendiger2017d, torres2021may}. On the other hand, the evaluation of  precision observables is a challenging task, primarily involving the numerical evaluation of higher-order perturbative cross sections and decay processes where ultraviolet and infrared divergences appear in intermediate steps.

A fully mathematical consistent regularization scheme that prevents the occurrence of symmetry breaking terms or spurious anomalies for the Standard Model and its extensions, and that is valid to all orders in perturbation theory is not available yet. In perturbative calculations at next-to-leading order (NLO) and beyond, infrared and ultraviolet divergences commonly arise due to the presence of loop diagrams and radiation from external legs in Feynman diagrams.  Finitude theorems guarantee the finiteness of perturbative calculations order by order under the hypothesis that a unitarity-preserving regularization method was employed to regulate these divergences. However, some regularization methods may not be completely consistent, especially when applied to extensions of the Standard Model or chiral theories.

For example, the cross section of single photon emission exhibits an infrared divergence in the limit of vanishing photon energy. This problem was initially solved by Bloch and Nordsieck (BN)~\cite{Bloch:1937pw}, who showed that the infrared divergence can be canceled out by considering inclusive processes where the bremsstrahlung contribution is combined with radiative corrections order by order in perturbation theory. Early in 1960, Kinoshita-Lee-Nauenberg (KLN) \cite{kinoshita1962mass, lee1964degenerate} stated that S-matrix elements squared are IR finite when a sum is performed
over final states and initial states within an energy window (degenerate states). In other words cancellation of IR divergences follows directly from
unitarity provided the measurement is inclusive enough: a hard parton can not be distinguished from a hard particle plus a
soft gluon or from two collinear partons with the same energy. Schematically, 
\begin{equation}
\text{KLN-theorem:} \sum_{i,f \in [E-\Delta,E+\Delta]} |\langle f|S|i\rangle |^2 = \text{finite}.
\end{equation}
In this way, the BN theorem is a special case of the KLN theorem.The computational challenge is to find the minimal set of diagrams needed for
IR finiteness.

At a certain order in perturbation theory, a subtle cancellation happens between IR divergences coming from coherently summed amplitudes (at the level of phase space integrals) and virtual IR (and possibly UV) divergent amplitudes.Therefore, a well-chosen regularization scheme is needed to handle the UV divergences, and also to ensure that the IR divergences cancel properly. Ideally, the regularization scheme must also respect the symmetries of the theory, and should not introduce spurious anomalies or breaking of symmetries.

 Standard (conventional) dimensional regularization (CDR) \cite{collins1984renormalization} and the closely related dimensional scheme of 't Hooft and Veltman (HV) \cite{veltman1972regularization} are the natural choice for Feynman amplitude calculations in gauge theories. Momenta and loop internal gauge fields are treated as $d$-dimensional objects, with $d=4-2\epsilon$ and $\epsilon \rightarrow 0$, while external gauge fields are $d$-dimensional in CDR and strictly four dimensional in HV. However dimensional extensions meet some challenges in the case of theories involving dimension specific quantities, such as the $\gamma_5$ matrix \cite{jegerlehner2001facts} and the Levi-Civita tensor in chiral and topological theories \cite{PhysRevD.58.125004}, as well as in  supersymmetric gauge theories \cite{STOCKINGER2006250}. In the latter, a mismatch between the number of degrees of
freedom of gauge fields (d) and gauginos (4)
breaks supersymmetry in CDR. 

Regarding $\gamma_5$ matrix issues in dimensional schemes, in the HV scheme, the loss of the anti-commuting property of $\gamma_5$ in d dimensions breaks BRST symmetry. The solution is the addition of   symmetry restoring counterterms (CT) using  the Breitenlohner-Maison (BM) \cite{breitenlohner1977dimensionally} scheme order by order in perturbation theory \cite{belusca2020dimensional,belusca2021two}.  A gauge invariant procedure has been advocated in \cite{tsai2011gauge, tsai2011maintaining} which allows to reduce the number of CT structures in the BM scheme, often denoted as "rightmost-position" method.  Some alternatives have also been considered to maintain the anti-commuting property of $\gamma_5$ in d dimensions whereas preserving gauge invariance and BRST symmetry at the cost of giving up the cyclic properties of the trace \cite{kreimer1990gamma5, kreimer1994role}~\footnote{See \cite{Chen:2023ulo} for recent subtleties when employing this scheme.}.
  
Apart from CDR and HV, other dimensional schemes have been developed to operate partially in the physical dimension. In such schemes, fields are treated differently with the help of additional metric spaces \cite{gnendiger2014infrared}, as in dimensional reduction (DRED) \cite{siegel1979supersymmetric, siegel1980inconsistency} and four dimensional helicity (FDH) \cite{bern1992computation, bern2002supersymmetric}. Both consider that gauge fields in the loop live in  quasi four dimensional spaces (Q4S) while external gauge fields are strictly four dimensional (4S) in FDH. DRED allows to consistently treat supersymmetric theories to 2-loop order \cite{stockinger2005regularization, signer2009using} and FDH  makes use of the efficient spinor helicity technique for the spin algebra of observables as these are defined in the physical dimension. This comes nevertheless at the cost of introducing extra fields that transform as scalars under Lorentz transformations, known as evanescent fields or $\epsilon$-scalars. They renormalize differently from the gauge field, and it is essential to kept track of these differences in order to have a scheme that respects unitarity~\cite{Kilgore:2012tb}. 

To fully explore the advantages of each of the schemes it is important that conversions between results in CDR, HV, FDH, and DRED  can be made at different steps of the calculation of a cross section. This can be done, for instance, using the scheme dependence of beta functions and anomalous dimensions \cite{Broggio:2015dga,Gnendiger:2016cpg}. 

On the other hand, methods that do not rely on dimensional extensions are emerging and being explored in diverse venues, with the intent of circumventing the above mentioned increase in complexity of dimensional schemes, such as the four dimensional unsubtracted (FDU) method~\cite{Hernandez-Pinto:2015ysa,Sborlini:2016hat}, the four dimensional regularization (FDR)~\cite{Pittau:2012zd}, and the implicit regularization (IReg)~\cite{Battistel:1998sz,BaetaScarpelli:2000zs,BaetaScarpelli:2001ix,cherchiglia2011systematic}. For recent views on some of these methods see \cite{aguilera2021stroll,arias2021brief}. One of the central objectives of these methods is to analytically implement a clear separation of UV and IR divergent content of Feynman amplitudes from the finite parts. The motivation is that this separation is valuable for simplifications in both the renormalization and the realization of the KLN theorem, while finite integrals are efficiently evaluated numerically. 

The FDU method is anchored on the loop-tree duality theorem and enables the cancellation of IR divergences at integrand level. In FDR the UV divergent content of an amplitude is isolated and discarded under certain conditions with the purpose of yielding directly a renormalized quantity at each order. IReg isolates the UV content in form of basic divergent integrals (BDI) and establishes all order relations among them which can be  conveniently used in the computation of renormalization functions (see section \ref{sec:ireg} for an overview). The prominent feature of BDIs is that they do not depend on masses and external momenta and endow a UV renormalization scale parameter. 

A direct set of conversion rules between non-dimensional and dimensional schemes at intermediate steps of a calculation is seen only up to NLO processes~\cite{gnendiger2017d}. Explicit calculations at NNLO processes and higher in IReg show that there are restrictions to a direct extension of transition rules~\cite{cherchiglia2021two}. Indeed this is also the case in a calculation at NNLO order comparing FDH and FDR showing that these schemes lack correspondence at intermediate steps. While the technical obstacle of transferring results between different schemes cannot be ignored, it is important to recognize that it may also present an opportunity for reordering calculational steps that could be advantageous in certain cases. Further investigation is required to fully comprehend the implications of such deviations in intermediate steps~\cite{torres2021may}.

 The purpose of this contribution is to use strong corrections to the vector-axialvector (V-A) decay $Z \rightarrow q \bar{q}$ and to the  charge neutral scalar (and pseudoscalar) decay $S  \rightarrow  q \bar{q}$ to NLO as a playground that encompasses different regularization subtleties: $\gamma_5$ matrix vertex, UV divergences, and virtual-real IR cancellation within the IReg framework. Such calculation lends
 insight in the relations among schemes involving the $\gamma_5$ matrix and IR finitude within IReg guaranteed by the KLN theorem. Moreover we make a direct comparison with CDR and DRED aiming at generalizing our strategy to physically more challenging and interesting setups beyond NLO.  

The paper is organized as follows. In section~\ref{sec:ireg} we present an overview of the IReg method. In section \ref{sec:decay} various processes involving scalar, pseudoscalar, pseudovector and vector decays are calculated and compared in section \ref{sec:compare} with well known results of CDR/HV  as well as with results of DRED/FDH and non-dimensional methods for the same processes. Conclusions summarize our results. We also present a set of appendixes. In the first, we perform an analysis of the role of the $\gamma_{5}$ matrix in IReg, justifying the usage of the rightmost approach when analyzing the decays  $Z \rightarrow q \bar{q}$, and $S  \rightarrow  q \bar{q}$. In the second, we discuss the delicate issue of observables containing an odd number of $\gamma_{5}$ matrices.

\section{Overview of Implicit Regularization}
\label{sec:ireg}
%%%%%%
In this section we present the rules of IReg focusing on one loop order and in the massless limit as for simplicity we consider decays into massless quarks (to study both soft and collinear infrared divergences). A complete $n$-loop set of rules can be found in  \cite{arias2021brief,cherchiglia2021two}.

In IReg, the extraction of the UV divergent content of a Feynman amplitude is done by using algebraic identities at the integrand level. This is done in alignment with Bogoliubov's recursion formula \cite{Bogoliubov:1957gp,Hepp:1966eg,Zimmermann:1969jj}, implying that the way the method defines an UV convergent integral respects locality, Lorentz invariance and unitarity~\cite{cherchiglia2011systematic}. IReg has been shown to respect abelian gauge invariance to $n$-loop order \cite{vieira:2015fra, ferreira:2011cv}, as well as non-abelian and SUSY symmetries in specific examples up to two-loop order \cite{cherchiglia2021two, cherchiglia:2021yxz, cherchiglia:2015vaa, Fargnoli:2010mf, Carneiro:2003id}. This is achieved in a constrained version of the method, in which surface terms (ST's), which are related to momentum routing of loops in Feynman diagrams, are set to zero. In the realm of applications, processes such as  $h\rightarrow \gamma\gamma$ \cite{Cherchiglia:2012zp}, $e^{-}e^{+}\rightarrow \gamma^{*} \rightarrow q\bar{q}(g)$ \cite{gnendiger2017d}, and $H\rightarrow gg (g)$\cite{pereira2023higgs} were studied at NLO.

Consider a general $1$-loop Feynman amplitude where we denote by $k$ the internal (loop) momenta, and $p_i$ the external momenta. To this amplitude, we apply the set of rules:

\begin{enumerate}
\item Perform Dirac algebra in the physical dimension. 

\item In order to respect numerator/denominator consistency,
as described in the reference~\cite{Bruque:2018bmy}, it is necessary to eliminate terms involving internal momenta squared in the numerator by dividing them out from the denominator. For instance,
\begin{equation}
\int_{k}  \dfrac{k^{2}}{k^{2}(k-p)^{2}}\bigg|_{\text{IREG}} \neq g^{\alpha\beta}\int_{k}  \dfrac{k_{\alpha}k_{\beta}}{k^{2}(k-p)^{2}}\bigg|_{\text{IREG}} \quad \mbox{where} \quad \int_k \equiv \int d^4k/(2 \pi)^4.
\end{equation}

\item Include a fictitious mass $\mu^{2}$ in all propagators, where the limit $\mu\rightarrow0$ must be taken at the end of the calculation. In the presence of IR divergences, a logarithm with $\mu^2$ will remain. Assuming that we have an implicit regulator, we apply the following identity in all propagators dependent on the external momenta $p_{i}$ 
	\begin{align}
	\frac{1}{(k-p_{i})^2-\mu^2}=\sum_{j=0}^{n-1}\frac{(-1)^{j}(p_{i}^2-2p_{i} \cdot k)^{j}}{(k^2-\mu^2)^{j+1}}
	+\frac{(-1)^{n}(p_{i}^2-2p_{i} \cdot k)^{n}}{(k^2-\mu^2)^{n}
		\left[(k-p_{i})^2-\mu^2\right]}.
	\label{ident}
	\end{align}
Here $n$ is chosen such that the UV divergent part only has propagators of the form $(k^{2}-\mu^{2})^{-j}$.
\item Express  UV divergences in terms of Basic Divergent Integrals (BDI's) of the form\footnote{UV divergences of quadratic nature (or higher) could also be kept in the framework of IReg, however, they will always cancel in theories that are multiplicative renormalizable. For some examples showing the explicit cancellation, see~\cite{BaetaScarpelli:2001ix,Sampaio:2005pc,Cherchiglia:2014gna,pereira2023higgs}.}
\bea
	I_{log}(\mu^2)&\equiv& \int_{k} \frac{1}{(k^2-\mu^2)^{2}},\quad \quad
    I_{log}^{\nu_{1} \cdots \nu_{2r}}(\mu^2)\equiv \int_k \frac{k^{\nu_1}\cdots
		k^{\nu_{2r}}}{(k^2-\mu^2)^{r+2}}.
\eea

\item Surface terms (weighted differences of loop integrals with the same degree of divergence) should be set to zero on the grounds of momentum routing invariance in the loop of Feynman diagrams. This constrained version automatically preserves gauge invariance:
\bea
\int_k\frac{\partial}{\partial k_{\mu}}\frac{k^{\nu}}{(k^{2}-\mu^{2})^{2}}&=&4\Bigg[\frac{g_{\mu\nu}}{4}I_{log}(\mu^2)-I_{log}^{\mu\nu}(\mu^2)\Bigg]=0.\label{ST1L}
\eea
 
\item A renormalization group scale can be introduced by disentangling the UV/IR behavior of BDI's under the limit $\mu\rightarrow0$. This is achieved by employing the identity
\beq
I_{log}(\mu^2) = I_{log}(\lambda^2) + \frac{i}{(4 \pi)^2} \ln \frac{\lambda^2}{\mu^2},
\label{SR1}
\eeq
It is possible to absorb the BDI's in the renormalisation constants (without explicit evaluation) \cite{Brito:2008zn}, and renormalisation functions can be readily computed using
\beq
\lambda^2\frac{\partial I_{log}(\lambda^2)}{\partial \lambda^2}= -\frac{i}{(4 \pi)^2}.
\eeq

\end{enumerate}

The above rules will be applied in the virtual contributions of the processes studied in this work. For the real contributions, we consider that the massless particles in the final state have the same fictitious mass $\mu$ introduced in step 3. This allows to also parametrise the infrared divergences coming from the real part in terms of logarithms of $\mu^2$, when performing the integration over a massive phase-space region. Regarding the matrix-element, it can still be computed in the massless limit, as we show in our examples.
 
Finally, since the treatment of the $\gamma_{5}$ matrix presents many subtleties, we will explain in appendix \ref{sec:gamma5} how the $\gamma_{5}$ can be consistently treated in connection with IReg.

\section{Decays to quarks and antiquarks within IReg}
\label{sec:decay}

In this section we present our main results. This work completes the study of NLO strong corrections to the decay of bosons to massless quarks and antiquarks within IReg. The off-shell photon decay was first considered in \cite{gnendiger2017d}, while here we will consider the decay of the Z-boson as well as of neutral scalars. Even though in the SM only one physical scalar particle is introduced (the Higgs boson), which is CP-even, for completeness we will also provide the result for the decay of a pseudo-scalar particle, which is present in many Beyond Standard Model extensions.

\subsection{NLO strong corrections to $Z\rightarrow q\bar{q}$}

In order to set our notation, we begin with the tree level decay rate of $Z\rightarrow q\bar{q}$. As standard, we need to compute
\begin{equation}\label{eq:Z1}
    \Gamma_t = \frac{1}{2m_{z}} \int\frac{d^3q}{(2\pi)^3 2 q_0} \frac{d^3\overline{q}}{(2\pi)^3 2 \overline{q}_0} \sum_{spin} |M_{\rm{tree}}|^2 (2\pi)^4 \delta^4(z-q-\overline{q}) \,,
\end{equation}
where $z^{\mu},m_z, q^{\mu}, \bar{q}^{\mu}$ are the four-momentum of the Z-boson, its mass, quark and anti-quark momenta, respectively, and $M_{\rm{tree}}$ is the tree-level amplitude 
\begin{equation}
    M_{\rm{tree}} =  \overline{u}(q) \cdot \frac{-ie \gamma^\mu Z_-}{\sin(2\omega)} \cdot v(\overline{q}) \epsilon_\mu (z)\,.
\end{equation}

We define $Z_\pm=(g_V \pm \gamma^5 g_A)$ with $g_V$ the vector component of the interaction, given by $g_V=I_3 - 2Q' sin^2(\omega)$, while $g_A$ is the axial component, given by $g_A=I_3$. $I_3$ is the third component of the particles' isospin, $Q'$ is the unitary charge and $\omega$ the weak mixing angle \cite{Romao:2012pq}. As can be noticed, there is a $\gamma_5$ matrix which may cause ambiguities under regularization. At the present stage, no particular treatment is required, since we are still at tree-level. After a straightforward calculation, one obtains\,\cite{novikov1999theory}
\begin{equation}
    \Gamma_t=\frac{e^2(g_V^2+g_A^2)m_{z}}{4 \pi \sin^2(2\omega)}.
\end{equation}

\subsubsection{Virtual decay rate}

\begin{figure}[h!]
\centering
\includegraphics[scale=0.3]{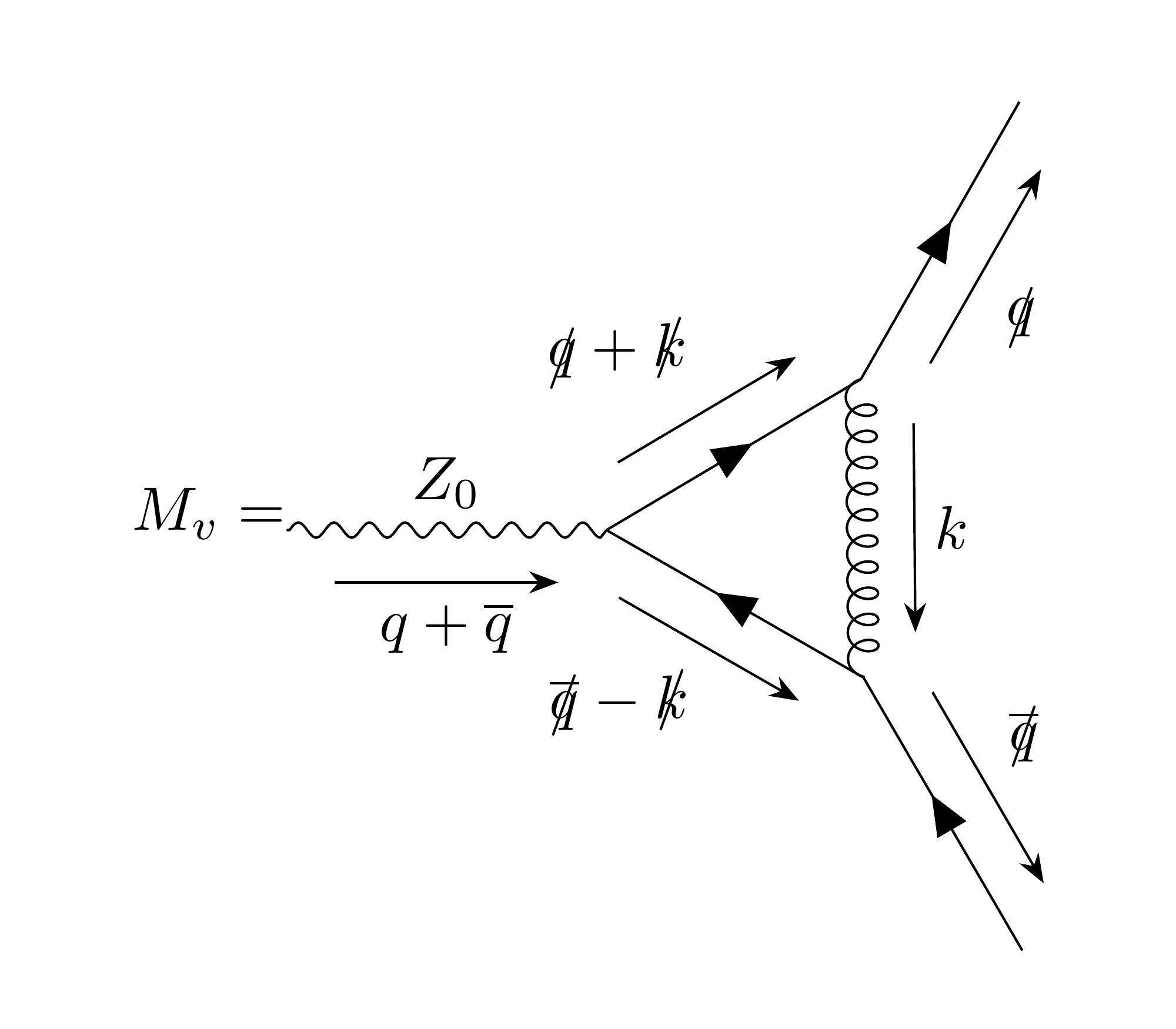}
\caption{Feynman diagram for the virtual contribution to decay $Z\rightarrow q\bar{q}$.}
\label{fig1}
\end{figure}

Regarding the NLO correction, we begin discussing the virtual decay rate, which stems from the diagram  of fig.\ref{fig1}.
%\begin{equation}
%    M_{v}=\feynmandiagram [baseline=(a.base), horizontal=a to b] {
%        a -- [photon, edge label=\(Z_0\), momentum'=\(q+\overline{q}\)] b,
%        f -- [fermion, reversed momentum'=\(\slashed{\overline{q}}\)] c -- [fermion, reversed momentum=\(\slashed{\overline{q}}-\slashed{k}\)] b -- [fermion, momentum=\(\slashed{q}+\slashed{k}\)] d -- [fermion, momentum'=\(\slashed{q}\)] e,
%        d -- [gluon, momentum=\(k\)] c,
%    };
%\end{equation}

We consider massless quarks, which renders the following amplitude

\begin{equation}\label{eq:z}
        M_v=\epsilon_\mu(z) \cdot \int_{k} \overline{u}(q) \cdot (-ig_{s}\gamma^\alpha t^a) \cdot \frac{-i}{\slashed{q}+\slashed{k}} \cdot \frac{-ie}{\sin(2\omega)} \gamma^\mu Z_{-} \cdot \frac{i}{\slashed{\overline{q}}-\slashed{k}} \cdot (-ig_{s}\gamma^\beta t^b) \cdot \frac{-ig_{\alpha \beta} \delta_{ab}}{k^2} \cdot v(\overline{q}) .
\end{equation}

Here $t^a$ are the color Gell-Mann matrices and $g_s$ the strong interaction constant (see also definitions below eq.\ref{eq:virtual_Z}). In order to deal with the $\gamma_5$ matrix, we will adopt the rightmost approach \cite{tsai2011gauge,tsai2011maintaining} which implies that $Z_\pm$ must be moved to the rightmost position. In the Appendix \ref{sec:gamma5} we justify the use of this procedure in the context of IReg for the processes calculated in the present work.  This allows us to rewrite the amplitude as follows
\begin{equation}\label{eq:Z}
    \begin{split}
        M_v=4 \frac{ g_{s}^2 (t^{a})^2 e}{\sin(2\omega)} \epsilon_\mu(z) \overline{u}(q) 
 [(q.\overline{q})I \gamma^\mu + (q^\mu - \overline{q}^\mu) \slashed{I} - \gamma^\mu (q - \overline{q})_{\alpha}  I^{\alpha} + \gamma_\delta I^{\delta \mu} -\frac{I_2}{2} \gamma^\mu] Z_{-} v(\overline{q}).
    \end{split}
\end{equation}
The integrals are regularized within IReg as below
\begin{align}\label{eq:int}
        I&=\int_{k} \frac{1}{k^2(q+k)^2(\overline{q}-k)^2} =\frac{b}{2 m_z^2}[\ln^2(\mu_0)+2i\pi \ln(\mu_0)-\pi^2],\\
        I^\mu&=\int_{k}  \frac{k^\mu}{k^2(q+k)^2(\overline{q}-k)^2} =\frac{b}{ m_z^2}(q^\mu-\overline{q}^\mu)\left[\ln\left(\mu_0\right) +i\pi+2\right],\\
        I^{\mu \nu}&=\int_{k} \frac{k^\mu k^\nu}{k^2(q+k)^2(\overline{q}-k)^2} 
         =\left[I_{log}(\mu^2)+b(\ln(\mu_0)+i\pi+3)\right]\frac{g^{\mu \nu}}{4}\nonumber\\&\quad\quad\quad\quad\quad\quad\quad\quad\quad\quad\quad\quad-\frac{b}{2m_z^2}\left[(q^\mu q^\nu+\overline{q}^\mu \overline{q}^\nu)(\ln(\mu_0)+2) + q^\mu \overline{q}^\nu + q^\nu \overline{q}^\mu\right],\\
        I_2&=\int_{k}  \frac{k^2}{k^2(q+k)^2(\overline{q}-k)^2} =I_{log}(\mu^2) + b[\ln(\mu_0)+i\pi+2],\label{eq:i2}
\end{align}
where $\mu_0=\mu^{2}/m_{z}^{2}$.

As can be seen, some of the integrals contain a UV divergence, $I_{log}(\mu^{2})$, that must still be removed by adopting a regularization scheme. We will choose the on-shell scheme, noticing that the Z-boson behaves as a spectator in our entire calculation. Thus, apart from factors such as $g_{V}, g_{A}$, and $\sin^{2}(2\omega)$ we will obtain a similar result to the process $e^{+}e^{-}\rightarrow \gamma^{*} \rightarrow q\overline{q}$. To be precise, in the on-shell scheme, the electromagnetic charge is renormalized by requiring $\overline{u}(p)\Gamma_{\mu}^{\gamma ee}(p,p) v(p) = ie\bar{u}(p)\gamma_{\mu}v(p) $, where $\Gamma_{\mu}^{\gamma ee}(p,p)$ is the amputated vertex function for $Ae\bar{e}$ and $e$, $A$ are the electron and photon fields respectively~\cite{Denner:1991kt}. Our calculation is completely analogous, yielding
\begin{align}
  M_v(z=0\;;\;q=\overline{q}) =  &\frac{e}{\sin(2\omega)}\epsilon_\mu(z) \int_{k} \overline{u}(q) \cdot (g_{s}\gamma^\alpha t^a) \cdot \frac{1}{\slashed{q}+\slashed{k}} \cdot  \gamma^\mu Z_{-} \cdot \frac{1}{\slashed{q}-\slashed{k}} \cdot (g_{s}\gamma_\alpha t^a) \cdot \frac{1}{k^2} \cdot v(q) \nonumber\\
  =&-\frac{ie}{\sin(2\omega)} \epsilon_\mu(z) \overline{u}(q)\gamma^{\mu}Z_{-} v(\overline{q})\left[C_{F}\frac{\alpha_{s}}{4\pi}\frac{I_{log}(\mu^{2})}{b}\right]\;.
\end{align}

Notice that the term left of the bracket is exactly the tree-level vertex $Zq\overline{q}$, and $e$ is a bare charge. By expressing it in terms of the renormalized charge $e=Z_{e}e_{r}=(1+\delta_{e})e_{r}$, it is possible to obtain the counterterm $\delta_{e}$ by
 \begin{equation}
     \delta_{e} = -C_{F}\left(\frac{\alpha_{s}}{\pi}\right)\frac{I_{log}(\mu^{2})}{4b}\;.
 \end{equation}
In section \ref{sec:scalar} we will present the renormalization function for the fermion field, $Z_{2}$, which fulfills $Z_{e}=Z_{2}$ as expected by virtue of the Ward identity.

Once the amplitude is regularized, it is straightforward to obtain the decay rate which, at NLO, is given by the interference term between the tree-level and one-loop amplitudes
\begin{equation}
    \Gamma_{v}=\frac{1}{16\pi m_z} [2Re(M_t^\dagger M_v)],
\end{equation}
with the result
\begin{equation}\label{eq:virtual_Z}
\Gamma_{v}=\Gamma_tC_{F}\left(\frac{ \alpha_s}{\pi}\right)\left[-\frac{\ln^2(\mu_0)}{2}-\frac{3}{2}\ln(\mu_0)-\frac{7}{2}+\frac{\pi^2}{2}\right].
\end{equation}
As standard, we used $(t^{a})^2=C_{F}$, and $\alpha_{s}=g_{s}^{2}/(4\pi)$. 

It is clear from the equation above that all dependence on the Z-boson vertex is included in $\Gamma_t$. Thus, the result above is compatible with the one obtained before for the process $e^{+}e^{-}\rightarrow \gamma^{*} \rightarrow q\bar{q}$ \cite{gnendiger2017d}, where the off-shell photon would play the same role as the Z-boson here.

\subsubsection{Real decay rate}

\begin{figure}[h!]
\centering
\includegraphics[scale=0.6]{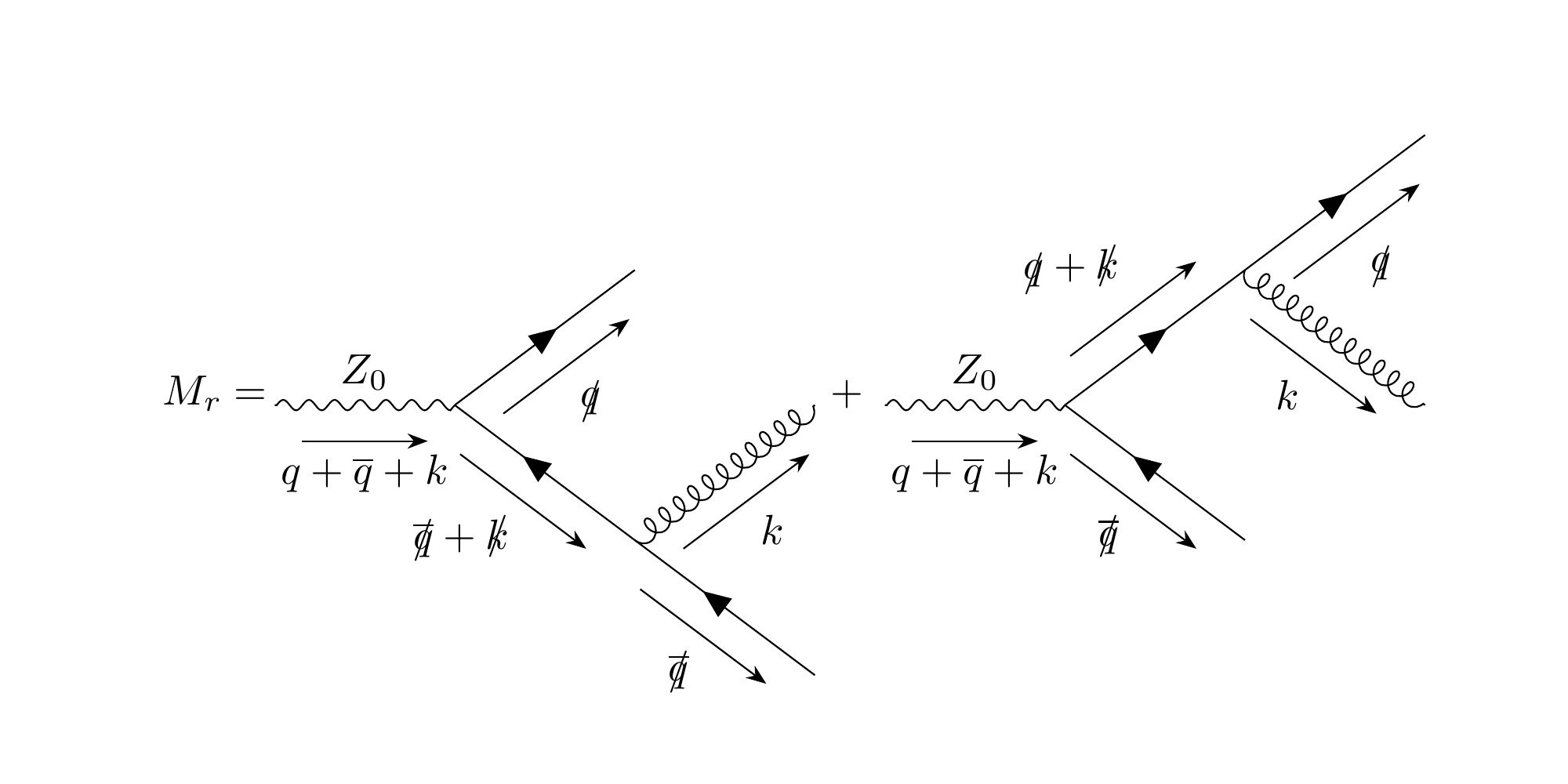}
\caption{Feynman diagrams for the real contribution to the decay $Z\rightarrow q\bar{q}$.}
\label{fig2}
\end{figure}

Once the virtual correction was obtained, we compute on this section the real contributions which are given by the diagrams of fig.\ref{fig2}. 

%\begin{equation}
%    M_{r}=\feynmandiagram [baseline=(a.base), horizontal=a to b, layered layout] {
%        a -- [photon, edge label=\(Z_0\), momentum'=\(q+\overline{q}+k\)] b,
%        b -- [fermion, momentum'=\(\slashed{q}\)] c,
%        b -- [anti fermion, momentum'=\(\slashed{\overline{q}}+\slashed{k}\)] d,
%        d -- [gluon, momentum'=\(k\)] e,
%        d -- [anti fermion, momentum'=\(\slashed{\overline{q}}\)] f,
%        {[same layer] c,d}
%    };
%    +
%    \feynmandiagram [baseline=(a.base), horizontal=a to b, layered layout] {
%        a -- [photon, edge label=\(Z_0\), momentum'=\(q+\overline{q}+k\)] b,
%        b -- [fermion, momentum=\(\slashed{q}+\slashed{k}\)] c,
%        c -- [fermion, momentum'=\(\slashed{q}\)] e,
%        c -- [gluon, momentum'=\(k\)] f,
%        b -- [anti fermion, momentum'=\(\slashed{\overline{q}}\)] d,
%        {[same layer] c,d}
%    };
%\end{equation}

The decay rate is obtained from the amplitude
\begin{equation}
        M_r=\epsilon_\mu (z) \overline{u}(q) \left[(-ig \gamma^\alpha t^a) \cdot \frac{-i}{\slashed{q}+\slashed{k}} \cdot \frac{-ie \gamma^\mu Z_-}{\sin(2\omega)} + \frac{-ie \gamma^\mu Z_-}{\sin(2\omega)} \cdot \frac{i}{\slashed{\overline{q}} + \slashed{k}} \cdot (-ig \gamma^\alpha t^a)\right] v(\overline{q}) \epsilon_\alpha (k).
\end{equation}

We introduce the following notation 
\begin{align}
         \chi&\equiv\frac{(z-q)^2}{m_z^2}-\frac{\mu^2}{m_{z}^2}=\frac{(\overline{q}+k)^2}{m_z^2}-\frac{\mu^2}{m_{z}^2},\\
         \overline{\chi}&\equiv\frac{(z-\overline{q})^2}{m_{z}^2}-\frac{\mu^2}{m_z^2}=\frac{(q+k)^2}{m_{z}^2}-\frac{\mu^2}{m_z^2}.
\end{align}
In terms of $\chi$, and $\overline{\chi}$, the modulus squared of the amplitude is given by

\begin{equation}
    \begin{split}
        |M_r|^2 
        = \frac{8 (t^a)^2 e^2 g^2 (g_V^2+g_A^2)}{\sin^2(2\omega) } \left[\frac{2-2\chi-2\overline{\chi}+\overline{\chi}^2+\chi^2}{(\chi+\mu_0) (\overline{\chi}+\mu_0)}\right].
    \end{split}
\end{equation}

We recall that we regularize the phase space integrals by introducing a fictitious mass $\mu$ in the propagator of the massless particles, which explains the presence of this term in $\chi$, and $\overline{\chi}$. To proceed we will make use of the results \cite{pittau2014qcd,gnendiger2017d}.
\begin{align}\label{eq:chi}
    &\int^{1-2\sqrt{\mu_0}}_{3\mu_0}\int^{\overline{\chi}^+}_{\overline{\chi}^-} \frac{1}{(\chi+\mu_0)(\overline{\chi}+\mu_0)} d\chi d\overline{\chi}=\frac{\ln^2(\mu_0)-\pi^2}{2}\nonumber\\
        &\int^{1-2\sqrt{\mu_0}}_{3\mu_0}\int^{\overline{\chi}^+}_{\overline{\chi}^-} \frac{\chi^a}{(\overline{\chi}+\mu_0)} d\chi d\overline{\chi}=
       -\frac{1}{a+1}\ln(\mu_0)-\frac{1}{a+1}\left(\frac{1}{a+1}+2\sum^{a+1}_{n=1} \frac{1}{n}\right) 
\end{align}
where $\overline{\chi}^\pm =  \frac{1-\chi}{2} \pm \sqrt{\frac{(\chi -3\mu_0)[(1-\chi)^2-4\mu_0]}{4(\chi+\mu_0)}} $. The same results hold if we replace $\chi\rightarrow\overline{\chi}$ and vice-versa in the integrand. 
Finally, the end result for the real contributions is

\begin{equation}\label{eq:real_Z}
    \Gamma_r=\Gamma_t C_{F}\left(\frac{ \alpha_{s}}{\pi}\right) \left[\frac{\ln^2(\mu_0)}{2}-\frac{\pi^2}{2}+\frac{\ln(\mu_0)}{2}+\frac{17}{4}\right],
\end{equation}
which is once again compatible with the result of \cite{gnendiger2017d}. 

Once the virtual and real contributions were calculated, we can obtain the NLO decay rate for the Z-boson to a pair of quark and antiquarks in the framework of IReg as

\begin{equation}
\Gamma_{\rm NLO}=\Gamma_t+\Gamma_v+\Gamma_r=\Gamma_t\left(1+\frac{3 C_{F} \alpha_s}{4\pi}\right)\;.
\end{equation}

Specializing to QCD, we have $C_{F}=4/3$, which renders the well-known result\,\cite{novikov1999theory}
\begin{equation}
\Gamma_{\rm NLO}=\Gamma_t\left(1+\frac{\alpha_s}{\pi}\right)\;.
\end{equation}

\subsection{NLO strong corrections to $S\rightarrow q\bar{q}$}

In this section we focus on the decay rate of a scalar (CP-even or odd) to a quark-antiquark pair. As before, we begin with the tree-level analysis. 

Similarly to the Z-boson decay case, eq.\eqref{eq:Z1}, we have the following decay rate for the scalar particle:

\begin{equation}
    \Gamma_{t}^{s} = \frac{1}{2m_{s}} \int \frac{d^3q}{(2\pi)^3 2 q_0} \frac{d^3\overline{q}}{(2\pi)^3 2 \overline{q}_0}\sum_{spin} |M_{\rm tree}^{s}|^2 (2\pi)^4 \delta^4(s-q-\overline{q})\,,
\end{equation}
where $s$ is the four-momentum of the scalar, and $m_{s}$ is its mass. By denoting the coupling of the scalar to the quarks by $\xi_{s}T$, where $T=\mathbb{1} (\gamma_5)$ for the CP-even (odd) scalar, it is straightforward to obtain the result
\begin{equation}
        \Gamma_{t}^{s} = \xi_s^2 \frac{m_{s}}{8\pi}\,.
\end{equation}
Once again, since we are at tree-level, no ambiguities related to the $\gamma_{5}$ matrix  occur. Notice that, in the context of the SM, $\xi_{s}\sim m_{q}$, where $m_{q}$ is the mass of the quark to which the scalar decays.\footnote{To be precise, in the SM we only have a CP-even scalar. However, the same reasoning applies for other BSM models, in which the coupling between the extra CP-odd boson is proportional to the quark mass.} Therefore, we cannot naively perform the NLO calculation for massless quarks, otherwise a null result would be  obtained. We will return to this point in the next section.
\\

\subsubsection{Virtual decay rate}
\label{sec:scalar}

\begin{figure}[h!]
\centering
\includegraphics[scale=0.3]{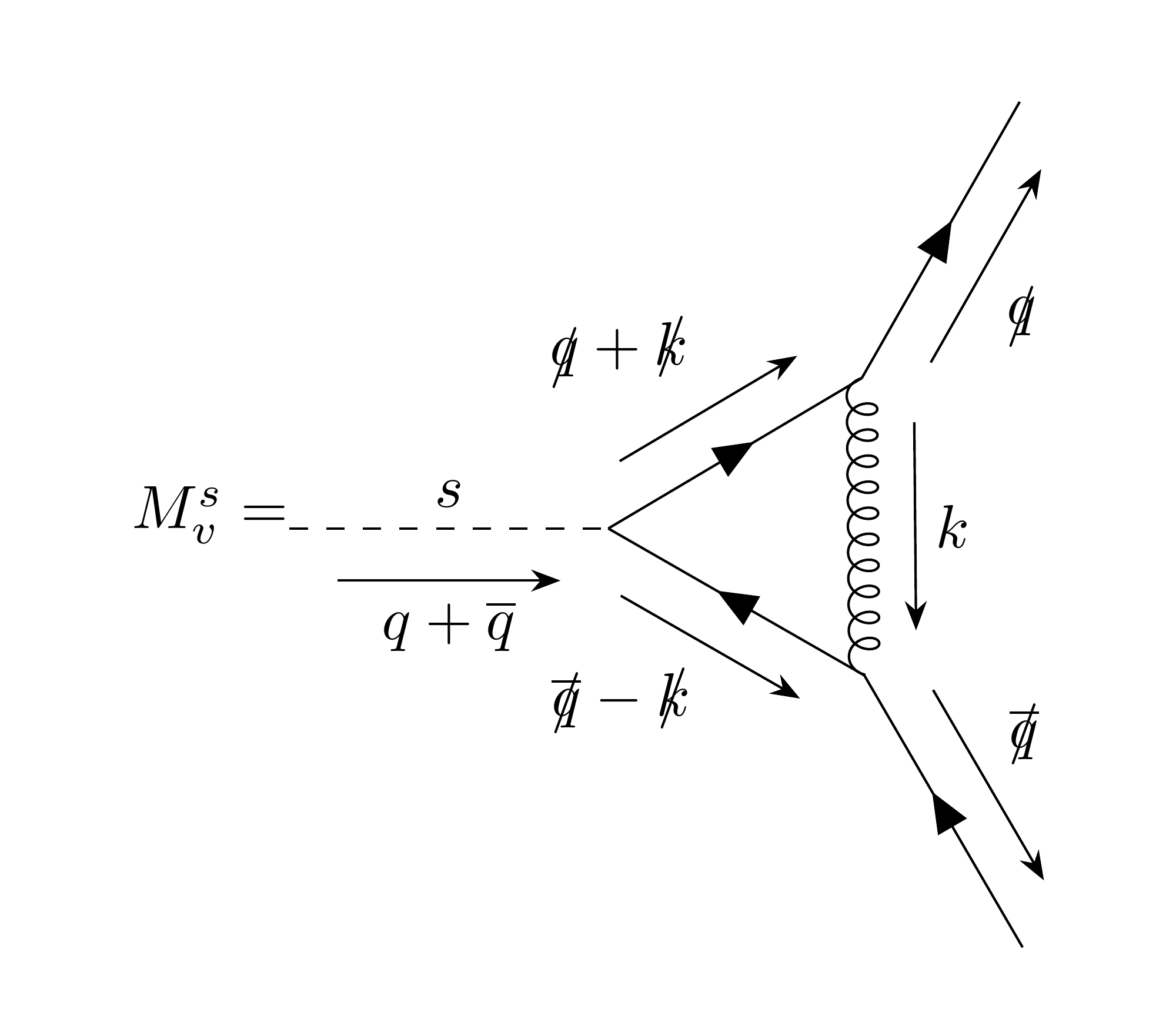}
\caption{Feynman diagram for the virtual contribution to decay $S\rightarrow q\bar{q}$.}
\label{fig3}
\end{figure}

Similarly to the Z-boson decay case, the virtual correction is due to the diagram of fig.\ref{fig3}.

%\begin{equation}
%    M_{v}^{s}=\feynmandiagram [baseline=(a.base), horizontal=a to b] {
%        a -- [scalar, edge label=\(s\), momentum'=\(q+\overline{q}\)] b,
%        f -- [fermion, reversed momentum'=\(\slashed{\overline{q}}\)] c -- [fermion, reversed momentum=\(\slashed{\overline{q}}-\slashed{k}\)] b -- [fermion, momentum=\(\slashed{q}+\slashed{k}\)] d -- [fermion, momentum'=\(\slashed{q}\)] e,
%        d -- [gluon, momentum=\(k\)] c,
%    };
%\end{equation}

For simplicity, we will treat only the case in which the mass of the quarks is arbitrarily small (massless limit), but the coupling $\xi_{s}$ is still non-null. In this case, we have the amplitude below:
\begin{equation}
        M_{v}^{s}=\int \frac{d^4k}{(2\pi)^4} \overline{u}(q) \cdot (-ig_{s}\gamma^\alpha t^a) \cdot \frac{-i}{\slashed{q}+\slashed{k}} \cdot (-i\xi_s T) \cdot \frac{i}{\slashed{\overline{q}}-\slashed{k}} \cdot (-ig_{s}\gamma^\beta t^b) \cdot \frac{-ig_{\alpha \beta} \delta_{ab}}{k^2} \cdot v(\overline{q}) \,.
\end{equation}

We recall that we are adopting the rightmost approach to deal with the $\gamma_{5}$ matrix, which allows us to write
\begin{equation}\label{eq:S}
        M_{v}^{s}= 4 g_{s}^2 (t^a)^2 \xi_s [(q\cdot\overline{q}) I + (\overline{q}_\mu-q_\mu) \cdot I^\mu - I_2] \overline{u}(q) T v(\overline{q})\,.
\end{equation}

The integrals in IReg are defined in eqs.\,(\ref{eq:int}-\ref{eq:i2}), with the end result
\begin{align}%\label{eq:virtual}
     M_{v}^{s}=  C_{F}\frac{\alpha_s}{\pi} \xi_s \left[-\frac{(\ln(\mu_0) +i\pi)^2}{4} + \frac{1}{b}I_{log}(\mu^2)\right] \overline{u}(q) T v(\overline{q})\,,
\end{align}
\noindent
where in this case $\mu_0=\mu^{2}/m_{s}^{2}$.

\begin{figure}[h!]
\centering
\includegraphics[scale=0.4]{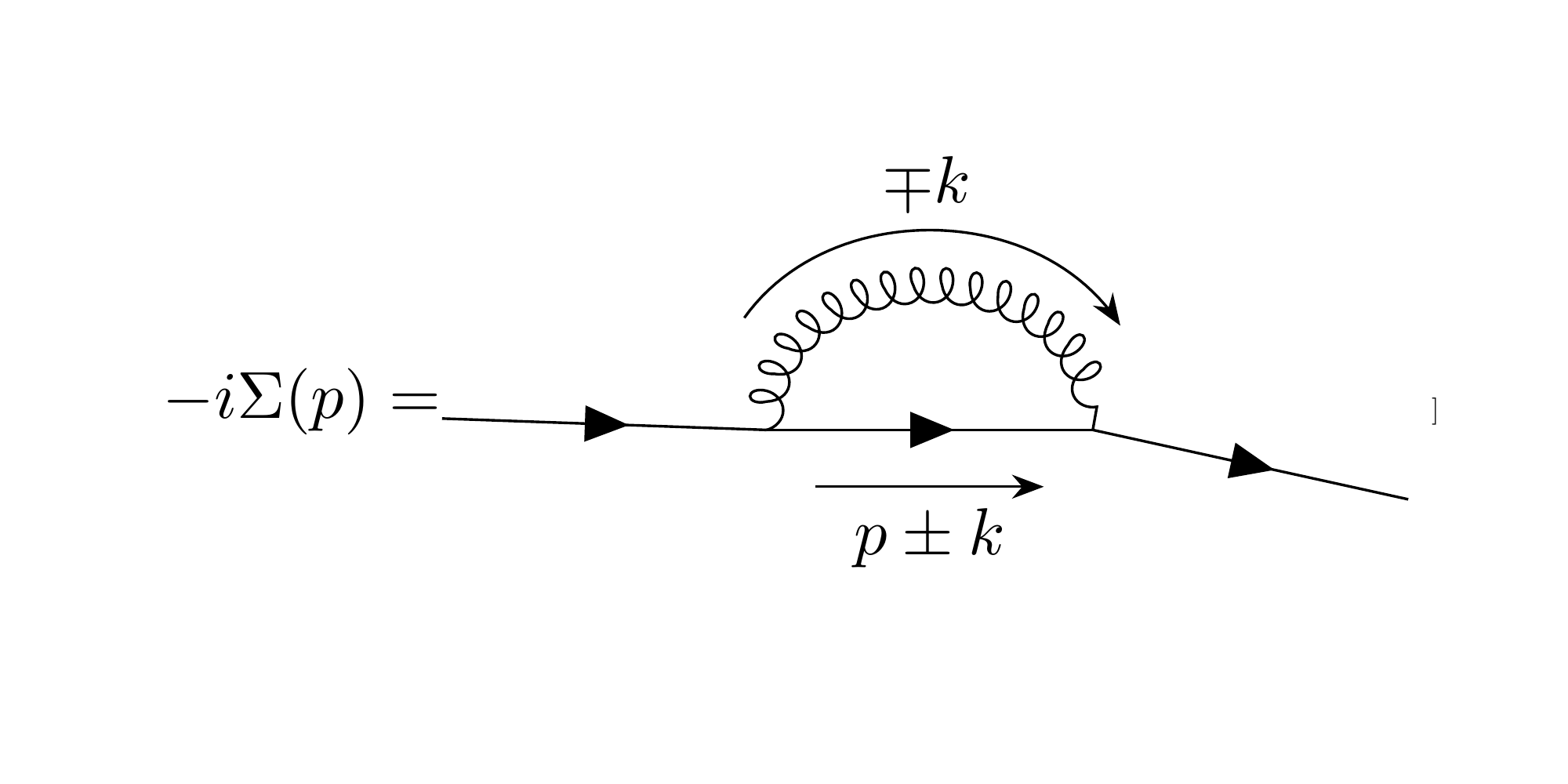}
\caption{Feynman diagram for the quark self-energy.}
\label{fig4}
\end{figure}

At this point, we would like to discuss some subtleties related to the massless limit we are adopting. As mentioned, for non-null $\xi_{s}$, we are implicitly assuming 
\begin{equation}\label{eq:bare}
    (\xi_{s})_{0} \sim (m_{q})_{0} = Z_{m_{q}}(m_{q})_{r}\,,  
\end{equation}
where $x_{0/r}$ denotes a bare/renormalized quantity, and $Z_{m_{q}}$ is the renormalization function of the quark mass. To obtain $Z_{m_{q}}$ in the context of the on-shell subtraction scheme, we will need to evaluate the strong corrections to the quark self-energy diagram which is given by fig.\ref{fig4}. 

%\begin{equation}
%    -i\Sigma(p)=\feynmandiagram [baseline=(a.base), horizontal=b to c] {
%        a -- [fermion] b,
%        b -- [fermion, momentum'=\(p \pm k\)] c,
%        c -- [fermion] d,
%        b -- [gluon, half left, momentum=\(\mp k\)] c
%    };
%\end{equation}

The on-shell renormalized mass is defined in such a way that it corresponds to the pole of the renormalized fermion propagator. In the context of IReg, we obtain
\begin{equation}
    Z_{m_{q}} = 1 + \delta_{m_{q}}= 1 - C_{F} \left(\frac{\alpha_{s}}{\pi}\right)\left\{\frac{5}{4} + \frac{3}{4}\left[\ln\left(\frac{\mu^2}{m_{q}^2}\right)+\frac{1}{b}I_{log}(\mu^2)\right]\right\}\,.
\end{equation}
We will also need the renormalization function of the fermion field which is given by\;\cite{sampaio2006implicit}
\begin{equation}
    Z_{2} = 1+ \delta_{2} = 1 -C_{F}\left(\frac{\alpha_{s}}{\pi}\right) 
        \frac{1}{4b}I_{log}(\mu^2)\,.
\end{equation}

Finally, we recall that, in order to obtain the decay rate, we need to compute
\begin{equation}
    \Gamma^{s} = \frac{1}{16\pi m_{s}} \sum_{spin}\left| M_{t}^{s} + M_{v}^{s}\right|^2\,.
\end{equation}
For consistency, $M_{t}^{s}$ must be expressed with the renormalized coupling $(\xi_s)_{r}$, and we need to account for the renormalization of the external fermion legs \cite{braaten1980higgs} which implies
\begin{equation}\label{eq:cons}
    \Gamma^{s} = \frac{m_{s}}{8\pi}(\xi_s)_{r}^{2} \left| 1+\delta_{m_{q}} + \delta_{2} + \frac{M_{v}^{s}}{(\xi_s)_{r}}\right|^2\,.
\end{equation}

By defining $\Gamma_{v}^{s}$ as the virtual contribution to the decay rate we are interested in, we obtain:
\begin{equation}
    \Gamma_{v}^{s} = \Gamma_{t}^{s}C_{F}\left(\frac{\alpha_{s}}{\pi}\right)\left\{-\frac{\ln^2(\mu_0)}{2}+\frac{\pi^2}{2} -\frac{5}{2} 
     -\frac{3}{2}\ln\left(\frac{\mu^2}{m_{q}^2}\right)
    \,.
    \right\}
\end{equation}

Notice that the UV integrals have canceled, as they should. The above result was obtained in the context of the on-shell renormalization scheme. In order to compare our result to dimensional schemes in the next section, we will translate it to the $\overline{MS}$ scheme of CDR, in which the relation below holds
\cite{neubert2020renormalization}:
\begin{equation}
m_{q}(\lambda)=m_{q}\left[1 - C_{F}\left(\frac{\alpha_{s}}{\pi}\right)\left(\frac{3}{4}\ln\left(\frac{\lambda^2}{m_{q}^2}\right)+1\right)\right]
\end{equation}
where $\lambda$ is a renormalization group scale. We then need to replace
\begin{equation}
    \xi_{s} \rightarrow \xi_{s}(\lambda)\left[1 + C_{F}\left(\frac{\alpha_{s}}{\pi}\right)\left(\frac{3}{4}\ln\left(\frac{\lambda^2}{m_{q}^2}\right)+1\right)\right]\,, 
\end{equation}
in eq.\ref{eq:cons}. The final result will be
\begin{equation}\label{eq:virtual}
    \Gamma_{v}^{s} = \Gamma_{t}^{s}C_{F}\left(\frac{\alpha_{s}}{\pi}\right)\left\{-\frac{\ln^2(\mu_0)}{2}+\frac{\pi^2}{2} -\frac{1}{2} 
     -\frac{3}{2}\ln\left(\mu_0\right)
        \right\}\,,
\end{equation}
where we have adopted $\lambda^2=m_{s}^{2}$.

\subsubsection{Real decay rate}

\begin{figure}[h!]
\centering
\includegraphics[scale=0.5]{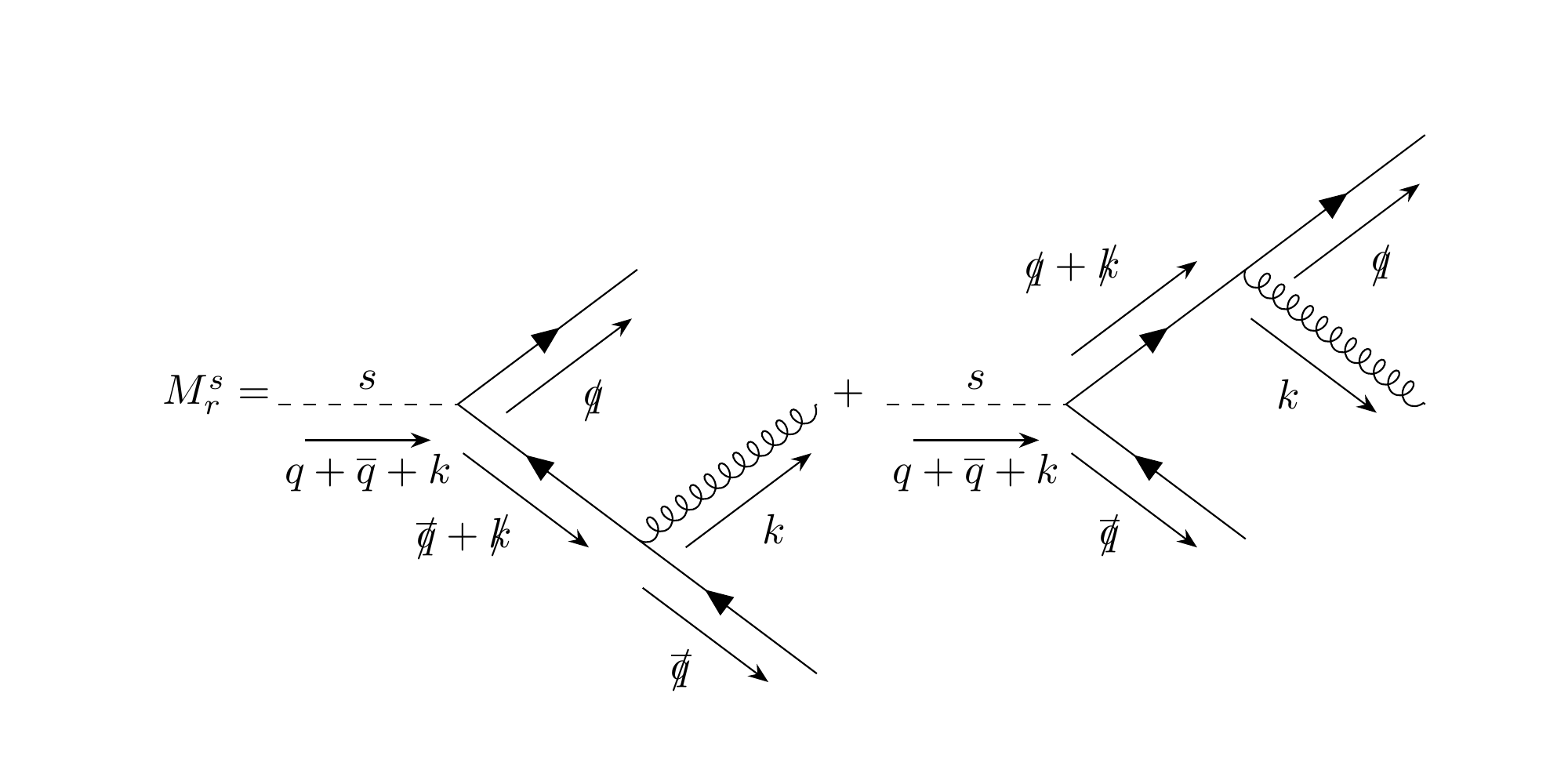}
\caption{Feynman diagrams for the real contribution to the decay $S\rightarrow q\bar{q}$.}
\label{fig5}
\end{figure}

The real contributions for the NLO strong correction to the decay $S\rightarrow q\bar{q}$ are given by the diagrams in fig.\ref{fig5}.

%\begin{equation}
%    M_{r}^{s}=\feynmandiagram [baseline=(a.base), horizontal=a to b, layered layout] {
%        a -- [scalar, edge label=\(s\), momentum'=\(q+\overline{q}+k\)] b,
%        b -- [fermion, momentum'=\(\slashed{q}\)] c,
%        b -- [anti fermion, momentum'=\(\slashed{\overline{q}}+\slashed{k}\)] d,
%        d -- [gluon, momentum'=\(k\)] e,
%        d -- [anti fermion, momentum'=\(\slashed{\overline{q}}\)] f,
%        {[same layer] c,d}
%    };
%    +
%    \feynmandiagram [baseline=(a.base), horizontal=a to b, layered layout] {
%        a -- [scalar, edge label=\(s\), momentum'=\(q+\overline{q}+k\)] b,
%        b -- [fermion, momentum=\(\slashed{q}+\slashed{k}\)] c,
%        c -- [fermion, momentum'=\(\slashed{q}\)] e,
%        c -- [gluon, momentum'=\(k\)] f,
%        b -- [anti fermion, momentum'=\(\slashed{\overline{q}}\)] d,
%        {[same layer] c,d}
%    };
%\end{equation}

Their amplitudes are easily obtained
\begin{equation}
    M_{r}^{s}=\overline{u}(q) \left[(-ig \gamma^\alpha t^a) \cdot \frac{-i}{\slashed{q}+\slashed{k}} \cdot (-i\xi_s T) -    
    (-i\xi_s T) \cdot \frac{i}{\slashed{\overline{q}} + \slashed{k}} \cdot (-ig \gamma^\alpha t^a)\right] v(\overline{q}) \epsilon^*_\alpha (k)\,,
\end{equation}
whose modulus squared is given by
\begin{equation}
        \left|M_{r}^{s}\right|^2= 16 g^2 \xi_s^2\; \left[\frac{2-2\chi_{s}-2\overline{\chi_{s}}+(\chi_{s}+\overline{\chi_{s}})^2}{(\chi_{s}+\mu_0)(\overline{\chi_{s}}+\mu_0)}\right]\;.
\end{equation}
For simplicity, we  introduced the notation
\begin{align}
         \chi_{s}&\equiv\frac{(s-q)^2}{m_{s}^2}-\frac{\mu^2}{m_{s}^2},\\
         \overline{\chi_{s}}&\equiv\frac{(s-\overline{q})^2}{m_{s}^2}-\frac{\mu^2}{m_{s}^2},
\end{align}
where $s=q+\overline{q}+k$. Using the results of eq.\eqref{eq:chi} together with
\begin{equation}
    \int^{1-2\sqrt{\mu_0}}_{3\mu_0}\int^{\overline{\chi}^+}_{\overline{\chi}^-}  d\chi d\overline{\chi}=\frac{1}{2}\;,
\end{equation}
we obtain the end result for the real decay rate

\begin{equation}\label{eq:real}
        \Gamma_{r}^{s}=\Gamma_{t}^{s}C_{F} \left(\frac{\alpha_s}{\pi} \right) \left[\frac{\ln^2(\mu_0)}{2}-\frac{\pi^2}{2}+\frac{3}{2}\ln(\mu_0)+\frac{19}{4}\right]\;. 
\end{equation}

Finally, adding the virtual, eq.\eqref{eq:virtual}, and the real, eq.\eqref{eq:real}, corrections we obtain the well-known result \cite{braaten1980higgs, sakai1980perturbative, inami1981renormalization, gorishnij1983width, djouadi2008anatomy} 
\begin{equation}\label{eq:final_scalar}
    \Gamma^{s} = \Gamma_{t}^{s} \left[1+C_{F}\frac{17 \alpha_s}{4\pi}\right]\,.
\end{equation}

\section{Comparison with  dimensional methods}
\label{sec:compare}

Once we have obtained the NLO results for the decay of a gauge boson or scalar to a pair of quark and antiquark in the framework of IReg, we aim to compare them to the results for the same processes obtained by using dimensional methods. As extensively discussed in~\cite{gnendiger2017d,torres2021may}, the Dimensional Reduction method (DRED) can be viewed as the most general of the dimensional schemes, allowing to reproduce the results in Conventional Dimensional Regularization (CDR), for instance, under certain limits. Moreover, in~\cite{torres2021may}, a detailed analysis of the decays $e^{-}e^{+}\rightarrow \gamma^{*} \rightarrow q\bar{q}$, and $h\rightarrow q\bar{q}$ were reviewed, both at NLO and NNLO. By adopting the $\gamma_5$ rightmost positioning approach, $\gamma_5$ matrices can be completely isolated out of the integrals, allowing us to directly use the results of \cite{torres2021may}. See, for instance, eqs.\ref{eq:Z},\ref{eq:S} in IReg which will have a completely analogous counterpart in dimensional schemes.

\subsection{NLO strong corrections to $S\rightarrow q\bar{q}$}

In the case of the scalar decay, there are no external gauge bosons present, implying that the treatment of the virtual corrections will require that only the internal gluon is split according to the DRED approach. We  denote the bare amplitude at NLO as  
\begin{equation}
    M_{\rm dim} = M_{\rm dim}^{(0)}\left[1+e^{-\epsilon\gamma_{E}}(4\pi)^{\epsilon}\left(\frac{\mu_{\rm dim}^{2}}{-m_{s}^{2}}\right)^{\epsilon}F^{(1)}_{\rm bare}\right]\,,
\end{equation}
where $\mu^{2}_{\rm dim}$ is the renormalization scale for dimensional methods, $F^{(1)}$ is a form factor, and $M_{\rm dim}^{(0)}=-i\overline{u}(q) \xi_s T  v(\overline{q})$ is the tree-level amplitude using our notation.

The form factor $F^{(1)}_{\rm bare}$ is given by \cite{torres2021may}
\begin{equation}
    F^{(1)}_{\rm bare} = C_{F}\frac{\alpha_{s}^{0}}{\pi}\left[-\frac{1}{2\epsilon^{2}} - \frac{1}{2} + \frac{\pi^{2}}{24}\right] + C_{F}\frac{\alpha_{e}^{0}}{\pi}n_{\epsilon}\left[\frac{1}{4\epsilon} + \frac{1}{2}\right]\,,
\end{equation}
where $n_{\epsilon}=2\epsilon$ and $\alpha_{e}$ is related to the coupling of the evanescent gluon to fermions. In the equation above, both couplings are to be considered bare, although, at NLO, the distinction will not be essential. On the other hand, since $\xi_{s}$ appear at tree-level, it is important to consider its renormalization
\begin{equation}
\xi^{0}_{s} = \xi_{s}\left[1+C_{F}\left(\frac{\alpha_s}{\pi}\right)e^{\epsilon\gamma_{E}}(4\pi)^{-\epsilon}\left(-\frac{3}{4\epsilon}-\frac{n_{\epsilon}}{8\epsilon}\right)\right]\,.
\end{equation}

It is straightforward to obtain the modulus squared of the amplitude, which yields the following decay rate
\begin{equation}
    \Gamma^{s}_{v (\rm dim)}=\Gamma^{s}_{t} C_F\Phi_2(\epsilon) c_\Gamma(\epsilon) m_{s}^{-\epsilon} \left[\frac{\alpha_s}{\pi}\left(-\frac{1}{\epsilon^2} - \frac{3}{2\epsilon} -1 + \frac{\pi^2}{2} +O(\epsilon) \right) + \frac{\alpha_\epsilon}{\pi}\left(\frac{n_\epsilon}{4\epsilon} + O(\epsilon)\right)\right]\,,
\end{equation}
where
\begin{equation}
          c_\Gamma (\epsilon) = (4\pi)^\epsilon \frac{\Gamma(1+\epsilon) \Gamma^2 (1-\epsilon)}{\Gamma (1-2\epsilon)} = 1 + O(\epsilon), \quad
    \Phi_2(\epsilon)=\left(\frac{4\pi}{m_{s}}\right)^\epsilon \frac{\Gamma(1-\epsilon)}{\Gamma(2-2\epsilon)}=1+O(\epsilon).
\end{equation}

It is valuable to make a comparison with the IReg result, as given by equation~\eqref{eq:virtual}. As previously noticed in \cite{gnendiger2017d}, the matching between the IR divergences in dimensional methods and IReg is given by $\frac{1}{\epsilon} \rightarrow \ln(\mu_0)$ and $\frac{1}{\epsilon^2} \rightarrow \frac{1}{2}\ln(\mu_0)^2$.

Moreover, by identifying $n_{\epsilon}=2\epsilon$ and setting $\alpha_{s}=\alpha_{e}$, we also recover the finite term. We notice that the result of CDR is obtained by setting $n_{\epsilon}=0$.

Regarding the real contribution, the decay rate is given by \cite{torres2021may}
\begin{equation}
    \Gamma^{s}_{r (\rm dim)}=\Gamma^{s}_{t} C_F\Phi_3(\epsilon) \left[\frac{\alpha_s}{\pi}\left(\frac{1}{\epsilon^2} + \frac{3}{2\epsilon} + \frac{21}{4} - \frac{\pi^2}{2} +O(\epsilon) \right) + \frac{\alpha_\epsilon}{\pi} \left(-\frac{n_\epsilon}{4\epsilon} + O(\epsilon^0)\right)\right]\,,
\end{equation}
where
\begin{equation}
    \Phi_3(\epsilon)=\left(\frac{4\pi}{h}\right)^{2\epsilon} \frac{1}{\Gamma(2-2\epsilon)}=1+O(\epsilon)\,.
\end{equation}

Notice that we may convert into the IReg result expressed by Eq.~\eqref{eq:real} under the same conditions imposed to virtual contributions. By adding both corrections we reproduce the well-known result expressed by Eq.~\eqref{eq:final_scalar}.

\subsection{NLO strong corrections to $Z\rightarrow q\bar{q}$}

In \cite{gnendiger2017d} the decay $e^{-}e^{+}\rightarrow q\bar{q}$ was computed at NLO using DRED. In that reference, only the strong correction was considered, which stands for modification only in the external legs containing quarks. Moreover, the mediator between leptons and quarks was an off-shell photon. The same process could occur with a Z-boson replaced mediator. If we are not interested in the initial states that will eventually generate the Z-boson, one can extract from the results of \cite{gnendiger2017d} the decay rate for the Z boson into a pair of quark and antiquark. In this case the virtual and real corrections are given by 
\begin{align}
    \Gamma_{v (\rm dim)} &= \Gamma_{t} C_F \Phi_2(\epsilon) c_\Gamma (\epsilon) m_{s}^{-\epsilon} \left[\frac{\alpha_s}{\pi}\left(-\frac{1}{\epsilon^2} - \frac{3}{2\epsilon} -4 + \frac{\pi^2}{2} +O(\epsilon) \right) + \frac{\alpha_\epsilon}{\pi}\left(\frac{n_\epsilon}{4\epsilon} + O(\epsilon^0)\right)\right]\,,\\
    \Gamma_{r (\rm dim)} &= \Gamma_{t}  C_f\Phi_3(\epsilon) \left[\frac{\alpha_s}{\pi}\left(\frac{1}{\epsilon^2} + \frac{3}{2\epsilon} + \frac{19}{4} - \frac{\pi^2}{2} +O(\epsilon) \right) + \frac{\alpha_\epsilon}{\pi} \left(-\frac{n_\epsilon}{4\epsilon} + O(\epsilon^0)\right)\right]\,.
\end{align}

As discussed before, the correspondence $\frac{1}{\epsilon} \rightarrow \ln(\mu_0)$ and $\frac{1}{\epsilon^2} \rightarrow {\frac{1}{2}}\ln^2(\mu_0)$ is verified and the IReg result is reproduced after setting $n_{\epsilon}=2\epsilon$ and identifying $\alpha_s=\alpha_e$ as seen in Eqns. \eqref{eq:virtual_Z} and \eqref{eq:real_Z}

\section{Concluding remarks}
\label{sec:conclusion}

To improve theoretical accuracy in precision observables, there has been considerable efforts to obtain beyond NLO corrections to processes probed at the LHC. It is generally acknowledged that the regularization of UV and IR poses challenges in automating higher order calculations. In response to this, alternative approaches have been developed that avoid partially or totally the use of dimensional continuation in the spacetime dimension such as IReg. These approaches offer the potential to simplify calculations or may be applicable in dimensional specific models such as chiral and topological ($\gamma_5$ matrix and Levi-Civita symbol issues) as well as supersymmetry models.

In this work, we studied at NLO the decay of  bosons (spin zero or one) into $q\bar{q}$ pairs in the framework of IReg, which is a fully quadridimensional regularization scheme. In particular, we computed the NLO strong corrections to the decay rates $\Gamma(Z\rightarrow q\bar{q})$ and $\Gamma(S\rightarrow q\bar{q})$, where $S$ can be a CP-even or odd scalar. We have verified that the KLN theorem is satisfied in our framework, and it is not necessary to introduce evanescent particles, unlike in partially dimensional methods such as FDH and DRED. We also compared IReg with these methods, showing that, regarding IR divergences, there is a precise matching rule between IReg and dimensional results at NLO, which was previously noticed in \cite{gnendiger2017d} regarding the process $e^{-}e^{+}\rightarrow \gamma^{*}\rightarrow q\bar{q}$ and later confirmed for the process $H\rightarrow gg$ as well \cite{pereira2023higgs}. Finally, since we considered a gauge boson with axial couplings, the presence of $\gamma_{5}$ matrices can potentially lead to  ambiguities in regularization and renormalization. To tackle this problem, we adopted the $\gamma_5$ rightmost position approach which is sufficient to render IReg a gauge invariant procedure in this case while reproducing the results obtained with more involved schemes in the literature. Therefore, IReg seems to be a feasible alternative to dimensional schemes, whose application and automation to NLO and beyond are active lines of research.

\section*{Acknowledgements}

We acknowledge support from Fundação para a Ciência e Tecnologia (FCT) through the projects
CERN /FIS-PAR /0040 /2019, CERN /FIS-COM /0035 /2019, UID /FIS /04564 /2020, and the grant FCT 2020.07172.BD. A.C.~acknowledges support from National Council for Scientific and Technological Development – CNPq through projects 166523\slash2020-8 and 201013\slash2022-3 and M. Sampaio acknowledges support from CNPq through grant 302790\slash2020-9.

\appendix

\section{Consistent treatment for $\gamma_{5}$}
\label{sec:gamma5}

In this appendix we present a more detailed analysis of the ambiguities that can be introduced by the treatment of the $\gamma_{5}$ matrix, and we show why the rightmost approach is sufficient in the examples presented.

As it is well-known, the $\gamma_{5}$ matrix is a strictly four dimension object, and it cannot be extended to $d$-dimensional. It means that, to treat chiral theories in dimensional schemes, one must first adopt a definition for $\gamma_{5}$, and check which of its four-dimensional properties are still fulfilled. Soon after the development of CDR, it was proposed in~\cite{breitenlohner1977dimensionally} to still define $\gamma_{5}$ in four-dimensions, while the remaining objects (Dirac matrices and momenta) are promoted to $d$-dimensions. This approach (BMHV scheme) necessarily invalidates the property $\{\gamma_{\mu},\gamma_{5}\}=0$, and requires a careful treatment of the sub-spaces in which CDR is defined. To be specific, CDR is defined in the space $QdS= 4S \oplus Q(-2\epsilon)S$, while $\gamma_{5}\in 4S$. Even though this approach breaks gauge invariance, it is the only alternative that delivers consistent (and unitarity-preserving) results at arbitrary loop order in CDR. 

In the case of methods defined in four-dimensions, somehow surprisingly, it was also shown that inconsistent results can be obtained in the presence of the $\gamma_{5}$ matrix within divergent integrals~\cite{porto2018bose,Bruque:2018bmy,Viglioni:2016nqc,Cherchiglia:2021uce}. In~\cite{Bruque:2018bmy} it was proposed a similar construction of dimensional schemes, in particular Dimensional Reduction, where Dirac matrices, with the exception of $\gamma_5$, are defined in a quasi-dimensional space $Q4S = QdS\oplus Q(2\epsilon)S$ \cite{stockinger2005regularization}. In contrast to DRED, where the momenta still need to be treated in $QdS$, in these methods momenta are also defined in $Q4S$, and we have the hierarchy $Q4S= 4S \oplus XS$. Here, $X$ is an auxiliary space, which does not need to be explicitly defined. Similar to the BHMV scheme, we obtain a consistent method at the price of not fulfilling some properties, $\{\gamma_{\mu},\gamma_{5}\}\neq0$, and breaking gauge invariance. 

In the specific case of IReg, the inconsistencies boil down to the contraction of internal momenta in Feynman amplitudes. To illustrate this point, consider the following results obtained in the framework of IReg

\begin{equation}
\int_{k}  \dfrac{k^{2}}{k^{2}(k-p)^{2}}=\int_{k}\dfrac{1}{(k-p)^{2}}=\lim_{\mu^{2}\rightarrow 0}\int_{k}  \dfrac{1}{(k-p)^{2}-\mu^{2}}=\lim_{\mu^{2}\rightarrow 0} \int_{k}  \dfrac{1}{k^{2}-\mu^{2}}=0
\label{Nk2}
\end{equation}
\begin{small}
\begin{equation}
\begin{aligned}
g^{\alpha\beta}\int_{k}  \dfrac{k_{\alpha}k_{\beta}}{k^{2}(k-p)^{2}}
&=g^{\alpha\beta}\left\{\left(\frac{p_{\alpha}p_{\beta}}{3}-\frac{g_{\alpha\beta}p^{2}}{12}\right)\left[I_{\text{log}}(\lambda^{2})- b \ln\left(-\frac{p^2}{\lambda^2}\right) + \frac{13b}{6}\right]-\frac{g_{\alpha\beta}b p^{2}}{24}\right\}\\
&=-\frac{b p^{2}}{6}
\label{gab_kakb}
\end{aligned}
\end{equation}
\end{small}

It is clear that if one insists to use $\{\gamma_{\mu},\gamma_{5}\}=0$, an ambiguity arises
\begin{align}\label{eq:first}
\int_{k}  \dfrac{\slashed{k}\gamma_{5}\slashed{k}}{k^{2}(k-p)^{2}}&
\overset{?}{=}
\frac{bp^{2}}{6}\gamma_{5},\quad\quad \mbox{using eq.\ref{gab_kakb} and $\{\gamma_{\mu},\gamma_{5}\}=0$,} \\
&\overset{?}{=}0, \quad\quad\quad\quad \mbox{using $\{\gamma_{\mu},\gamma_{5}\}=0$ and eq.\ref{Nk2}.}\label{eq:second}
\end{align}

In order to avoid these ambiguities, one defines
\begin{equation}
\gamma_{5}=-\frac{i}{4!}\epsilon_{abcd}\bar{\gamma}^{a}\bar{\gamma}^{b}\bar{\gamma}^{c}\bar{\gamma}^{d}
\label{eq:def G5}    
\end{equation}
where we use a $\bar{}\;$ to denote an object pertaining to $4S$. Since the Dirac matrices are defined in $Q4S$, we have the properties
\begin{align}
\{\gamma_{\mu},\gamma_{\nu}\}&=2g_{\mu\nu}\mathbb{1}; \quad \{\bar{\gamma}_{\mu},\bar{\gamma}_{\nu}\}=\{\gamma_{\mu},\bar{\gamma}_{\nu}\}=2\bar{g}_{\mu\nu}\mathbb{1}; \quad \gamma_{\mu}\gamma^{\mu}=\gamma_{\mu}\bar{\gamma}^{\mu}=4\;\mathbb{1}\\
\{\bar{\gamma}_{\mu},\hat{\gamma}_{\nu}\}&=0; \quad \{\gamma_{\mu},\hat{\gamma}_{\nu}\}=\{\hat{\gamma}_{\mu},\hat{\gamma}_{\nu}\}=2\hat{g}_{\mu\nu}\mathbb{1};\quad \gamma_{\mu}\hat{\gamma}^{\mu}=\bar{\gamma}_{\mu}\hat{\gamma}^{\mu}=\hat{\gamma}_{\mu}\hat{\gamma}^{\mu}=0.
\end{align}
\begin{align}
\{\bar{\gamma}_{\mu},\gamma_{5}\}=0; \quad \{\gamma_{\mu},\gamma_{5}\}=2\gamma_{5}\hat{\gamma_{\mu}}; \quad [\hat{\gamma}_{\mu},\gamma_{5}]=0
\end{align}
where we denoted by $\hat{}\;$  an object belonging to the $X$ space. In view of the above properties, the previous integral is given by~\cite{Cherchiglia:2021uce}

\begin{align}
\int_{k}  \dfrac{\slashed{k}\gamma_{5}\slashed{k}}{k^{2}(k-p)^{2}}&=
2\gamma_{5}\int_{k}  \dfrac{\hat{\slashed{k}}\slashed{k}}{k^{2}(k-p)^{2}}-\int_{k}  \dfrac{\gamma_{5}k^2}{k^{2}(k-p)^{2}}\nonumber\\
&=2\gamma_{5}\int_{k}  \dfrac{k^2}{k^{2}(k-p)^{2}}-2\gamma_{5}\int_{k}  \dfrac{\bar{k}^2}{k^{2}(k-p)^{2}}\nonumber\\
&=-2\bar{g}_{ab}\gamma_{5}\int_{k}  \dfrac{\bar{k}^{a}\bar{k}^{b}}{k^{2}(k-p)^{2}}=\frac{bp^{2}}{3}\gamma_{5}. \label{eq:third}
\end{align}

After laying down the main ideas, we tackle the specific examples studied in this work. For ease of the reader, we repeat below the virtual contribution to the decay $Z\rightarrow q\bar{q}$
\begin{equation}
        M_v=\epsilon_\mu(z) \cdot \int_{k} \overline{u}(q) \cdot (-ig_{s}\gamma^\alpha t^a) \cdot \frac{-i}{\slashed{q}+\slashed{k}} \cdot \frac{-ie}{sin(2\omega)} \gamma^\mu (g_V - \gamma^5 g_A) \cdot \frac{i}{\slashed{q}_{b}-\slashed{k}} \cdot (-ig_{s}\gamma^\beta t^b) \cdot \frac{-ig_{\alpha \beta} \delta_{ab}}{k^2} \cdot v(\overline{q})\,.
\end{equation}
where we redefined the momentum of the antiquark by $q_{b}$ to avoid confusion.

The only possibly ambiguous part is proportional to $g_{A}$. Focusing only on this part, we can perform an analogous computation to eq.~\ref{eq:third}. One finds that on top of the result already found using the rightmost approach (eq. \ref{eq:z}), we will have terms of the form
\begin{equation}
   M_v \supset \int_{k} \frac{(k^{2} - \overline{k}^{2})\gamma_{\mu}\gamma_{5}}{k^{2}(k+q) ^{2}(k-q_{b})^{2}}\,, \int_{k} \frac{(\overline{k}^{\mu} - k^{\mu} )\slashed{\overline{k}}\gamma_{5}}{k^{2}(k+q) ^{2}(k-q_{b})^{2}}\,,\int_{k} \frac{k^{\mu}(\slashed{k}-\slashed{\overline{k}} )\gamma_{5}}{k^{2}(k+q) ^{2}(k-q_{b})^{2}}
\end{equation}

However, by close inspection of the on-shell result of these integrals in the framework of IReg (eqs. \ref{eq:int}-\ref{eq:i2}), we find, for instance,
\begin{equation}
    \int_{k} \frac{(k^{2} - \overline{k}^{2})}{k^{2}(k+q) ^{2}(k-q_{b})^{2}} = \int_{k} \frac{k^{2} }{k^{2}(k+q) ^{2}(k-q_{b})^{2}}  - \overline{g}_{ab}\int_{k} \frac{ k^{a}k^{b}}{k^{2}(k+q) ^{2}(k-q_{b})^{2}} = 0\,.
\end{equation}
A similar result holds for the other integrals. Therefore, for the particular example of the decay $Z\rightarrow q\bar{q}$ at NLO, we find that there is no extra term arising from consistently treating the $\gamma_{5}$ matrix in IReg. This justifies the usage of the rightmost approach in our calculation. A complete analogous reasoning can be applied to the decay  $S\rightarrow q\bar{q}$ at NLO.

%%%%%%%%%%%%%%%%%%%%%%%%%%%%%%Note on the chiral anomaly%%%%%%%%%%%%%%%%%%%%%%%%%%%%%%%%%%%%%%%
\section{Note on the chiral anomaly}

The processes analyzed so far dealt with the $\gamma_5$ matrix placed in an open fermionic line (Z and pseudoscalar amplitudes), for which we were able to show that the rightmost method is compatible with the  approach devised for IReg mentioned in the appendix~\ref{sec:gamma5}. For closed fermionic lines, the occurrence of the $\gamma_5$ matrix must be treated in connection with the trace properties of the Dirac algebra within divergent integrals. As mentioned in the introduction cyclic properties of the trace may or may not be kept, depending on the method applied. 
To illustrate the procedure within IReg we very shortly review the diagrammatic calculation of the chiral anomaly  in the context of the anomalous pion decay in two photons, $\pi^0\rightarrow\gamma\gamma$, proceeding through one-loop quark diagrams, where a trace must be taken over an odd number of $\gamma_5$ matrices occurring in the amplitude and related Adler-Bardeen-Bell-Jackiw (ABJ) anomaly~\cite{Bell:1969ts,Adler:1969gk}.  

\begin{figure}[h!]
\centering
\includegraphics[trim=0mm 100mm 0mm 100mm,scale=0.5]{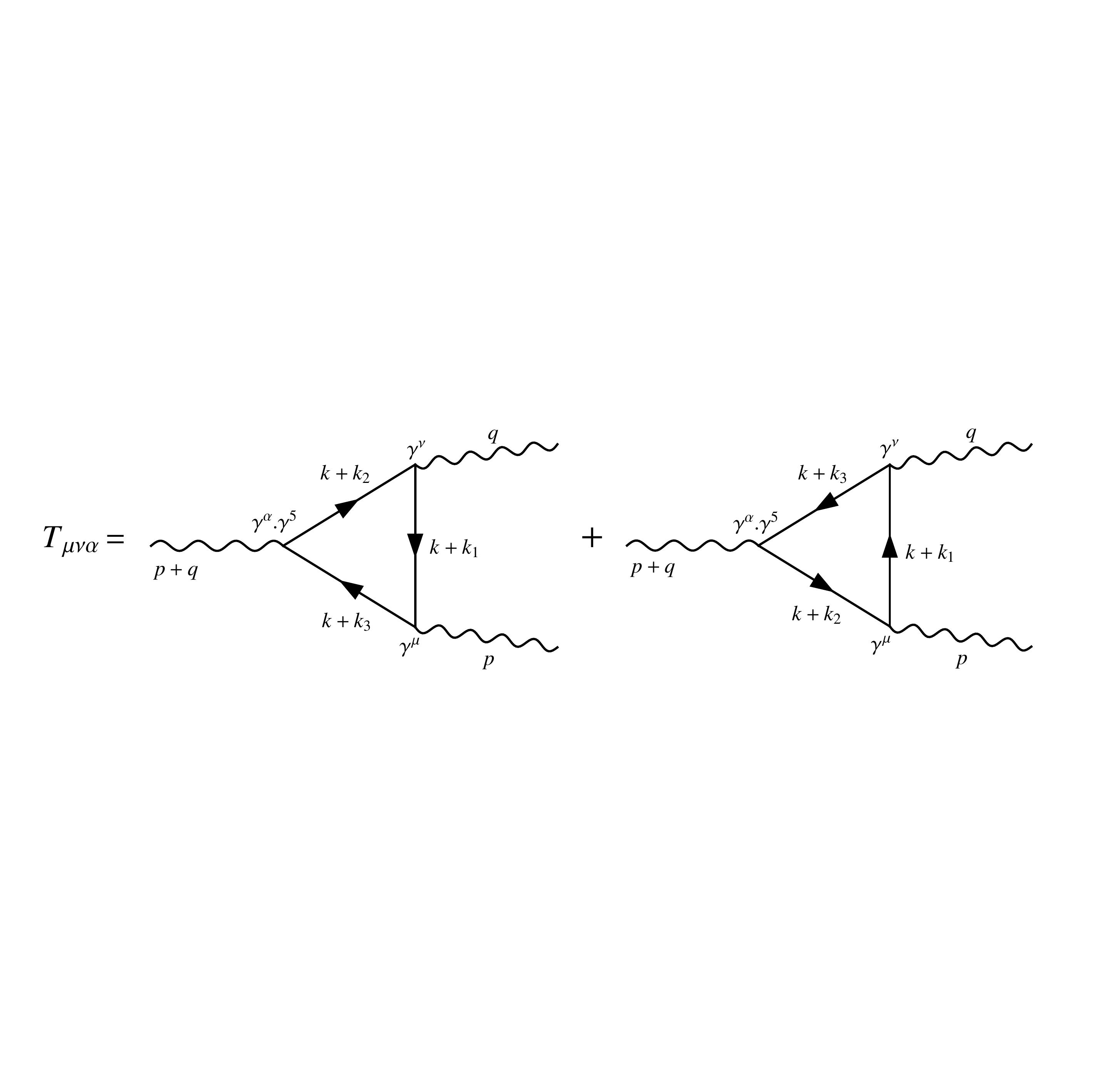}
\caption{Triangle diagrams which contribute to the ABJ anomaly. Following Ref.~\cite{Viglioni:2016nqc}, the internal lines are labeled with arbitrary momentum routing.}
\label{triangle}
\end{figure}

In IReg the cyclic property of the trace is maintained, therefore the rightmost position method results in using the anti-commutator to place the $\gamma_5$ matrix in any position within the trace. This turns out not to be sufficient to secure the anomaly in the pertaining Ward identity. In other words, although the value of the anomaly is correctly reproduced, it appears in the vectorial instead of axial Ward identity (WI), when surface terms (ST) are set to zero. If one insists that gauge invariance should result from setting ST to zero, the problem is resolved within IReg by recurring to the symmetrization of the trace, which is implemented by taking the definition $\gamma_5=\frac{1}{4!} \epsilon_{\mu\nu\alpha\beta} \gamma^\mu \gamma^\nu \gamma^\alpha \gamma^\beta$.  These issues have been thoroughly addressed in several works of IReg, see e.g.~\cite{porto2018bose,Bruque:2018bmy,Viglioni:2016nqc,Cherchiglia:2021uce}.  
With this rule the direct calculation of the pertinent quark triangle Feynman diagrams in IReg, see Figure~\ref{triangle}
, yields for the axial vector vector ($AVV$) amplitude
\begin{eqnarray}
\label{AVV}
T^{\mu\nu\alpha}=-\frac{1}{4\pi^2} (1+a)\epsilon^{\mu\nu\alpha\beta} (q-p)_\beta +T^{\mu\nu\alpha}_{fin}   \nonumber \\
\end{eqnarray}
 and  the following  vectorial  and axial Ward identities 
\begin{eqnarray}
\label{WI}
&&p_\mu T^{\mu\nu\alpha}=-\frac{1}{4\pi^2} (1+a)\epsilon^{\alpha\nu\beta\lambda}p_\beta q_\lambda   \nonumber \\
&&q_\nu T^{\mu\nu\alpha}=-\frac{1}{4\pi^2} (1+a)\epsilon^{\alpha\mu\beta\lambda}p_\beta q_\lambda \nonumber \\
&& l_\alpha T^{\mu\nu\alpha}=2 m T_5^{\mu\nu} +\frac{1}{2\pi^2} a \epsilon^{\mu\nu\beta\lambda}p_\beta q_\lambda. \
\end{eqnarray}
with $l=p+q$.
Here $T^{\mu\nu\alpha}_{fin} $ is the part of the amplitude that contains only finite integrals after an arbitrary valued surface term, denoted by $1+a$, has been isolated, signalizing that the WI can not be simultaneously satisfied. Choosing  $a= - 1$   gauge invariance is ensured.  In operator language the axial vector current density $ j^5_\mu(x)=\bar {\psi}(x) \gamma_\mu \gamma_5 \psi(x)$ acquires then the quantum correction 
\begin{equation}
\label{jbare}
\partial^\mu j^5_\mu(x)=2 i m j^5 - \frac{1}{16 \pi^2} \epsilon^{\mu\nu\alpha\beta}F_{\mu\nu}F_{\alpha\beta}
\end{equation}
where $ j^5(x)=\bar {\psi}(x) \gamma_5 \psi(x)$ is the pseudoscalar current density present at classical level
and   $F^{\mu\nu}$ denotes the electromagnetic field strength. These results are a stringent test of IReg. 
The  $T_5^{\mu\nu}$  represents the pseudoscalar vector vector (PVV) amplitude and only involves UV finite integrals after evaluation of the Dirac trace, being therefore regularization independent.

The remaining discussion relating these WI to the pion radiative decay is beautifully addressed in many text books, for instance \cite{bertlmann2000anomalies, itzykson2012quantum, cheng1994gauge, peskin1995introduction}. Here we use it to show that once the WI have been correctly identified within IReg, the model independent aspects of the radiative decay must follow as well.
The above WI are of special relevance to  particles of Goldstone nature as the pion. The axial current acquires dynamical significance through the partial conservation of the axial current approximation (PCAC), which states that upon spontaneous chiral symmetry breaking  the   axial symmetry current $A_\mu^{5,a}$ of the hadronic model has non-vanishing matrix elements between the vacuum and a Goldstate state and is conserved in the Goldstone limit $m_\pi^2=0$, $<0|\partial^\mu A_\mu^{5,a} (x) |\pi^b,p> = i f_\pi m_\pi^2 e^{-i p.x} \delta_{ab} $, where $f_\pi \sim 93$ MeV is the weak decay constant of the pion and $a,b$ denote isospin components. However in the presence of electromagnetic gauge fields it must be modified to accommodate the anomaly contribution 
\begin{equation}
\label{mod}
\partial^\mu A_\mu^{5,3} = f_\pi m_\pi^2 \pi^0 -\frac{\alpha}{8\pi} \epsilon_{\mu\nu\alpha\beta} F^{\mu\nu}F^{\alpha\beta}, 
\end{equation}
with $\alpha=\frac{e^2}{4\pi}$, $\pi^0$ the neutral pion field and the index $3$ stands for the third component of the axial vector current  %$j_\mu^{5,a}(x)=\bar {\psi}(x) \gamma_\mu \gamma_5 \frac{\tau^a}{2} \psi(x)$. %
The first term contains information about the pseudo-Goldstone nature of the pion and the second is the anomalous divergence of the axial vector current.% It carries an extra factor $\frac{1}{2}$ as compared to (\ref{jbare}) since the axial current has this factor associated with the Pauli matrix.

The modification of the PCAC relation can be sketched as follows. We start by considering  $T_5^{\mu\nu}$ in eq. (\ref{WI}). It has the same Lorentz structure as the $\pi^0\rightarrow\gamma\gamma$ amplitude in chiral quark models of the pion with  $g_\pi (\bar q i \gamma_5 \tau_a \pi^a q)$ effective Lagrangian description, such as in the $SU(2)_A\times SU(2)_V$ linear sigma model (LSM) with quark degrees of freedom, therefore it is convenient to adopt the LSM for the present discussion. The hadronic coupling $g_\pi$ is given by the  celebrated Goldberger-Treiman (GT) relation $g_{\pi}=\frac{m g_A}{f_\pi}$ in the limit $l^2\rightarrow 0$, where $g_A$ is associated to the neutron beta decay (set here to  $g_A=1$ \cite{itzykson2012quantum}) and $m$ is the constituent quark mass. The Noether current $A_\mu^{5,3}= \bar {q} \gamma_\mu \gamma_5 \frac{\tau^3}{2} q +$  bosonic contributions. 
The $T_5^{\mu\nu}$ amplitude can be readily converted to the physical radiative decay amplitude of the pion, after taking into account isospin and $N_c$ color factors appropriately.  A factor of 2 also arises associated with the LSM coupling $\tau_i$ as compared to $\tau_3/2$ in the fermionic current. 
Direct evaluation yields in the Goldstone limit for the $\pi^0\rightarrow\gamma\gamma$ amplitude
\begin{eqnarray}
\label{PiVV}
&&T_\pi^{\mu\nu}=g_\pi N_c 2 Tr[\frac{1}{2} \tau_3 \{Q,Q\}] T_5^{\mu\nu}  \\
&&T_5^{\mu\nu}=\frac{\alpha}{m}\frac{1}{\pi} \epsilon_{\mu\nu\alpha\beta} p^\alpha q^\beta \
\end{eqnarray}
where $Q=\frac{1}{2}(\frac{1}{3}+\tau_3)$ is the quark charge matrix at the electromagnetic vertex, the trace is over isospin, and a smooth behavior as $l^2\rightarrow 0$  is implied. With $h= N_c Tr[\frac{1}{2} \tau_3 \{Q,Q\}]=\frac{1}{2}$ one obtains $T_\pi^{\mu\nu}=\frac{\alpha}{ \pi f_\pi} \epsilon_{\mu\nu\alpha\beta} q^\alpha p^\beta$.

On the other hand the pion  field in the evaluation of the quark loop coupling to photons can be reduced using the Lehmann, Symanzik, Zimmermann (LSZ) reduction formula \cite{lehmann1955formulierung} yielding after using the modified PCAC  relation eq. (\ref{mod})
\begin{equation} 
\label{LSZ}
l_\alpha {\bar T}^{\mu\nu\alpha}(l^2)=\frac{f_\pi m_\pi^2}{(m_\pi^2-l^2) } T_\pi^{\mu\nu}(l^2) -\frac{\alpha}{\pi} \epsilon^{\mu\nu\beta\lambda} p_\beta q_\lambda.
\end{equation}
where ${\bar T}^{\mu\nu\alpha}$ carries the isospin and color factor $h$ defined above.
The first term on the right hand side stems from the original PCAC relation. In the absence of the anomalous contribution  the conservation of the axial vector current  as  $l^2\rightarrow0$ can only be achieved if lim $T_\pi^{\mu\nu}(0) =0$, as observed by Sutherland \cite{sutherland1967current} and Veltman \cite{veltman1967theoretical}, which would be in contradiction with the explicit evaluation of the amplitude.
From eq. (\ref{LSZ}) one obtains finally that the model independent leading contribution to the radiative pion amplitude   $T_\pi^{\mu\nu}(0)$ is  given by the anomalous term  and coincides with the expression eq. (\ref{PiVV}). 

To summarize, although IReg operates in the physical dimension, processes involving $\gamma_5$ still require some care, as explained in the Appendix~\ref{sec:gamma5}. We have found at one loop level that in open fermionic lines the $\gamma_5$ can be treated as in the rightmost position method,  where it takes a spectator role and WI are preserved. In closed fermionic lines involving an odd number of $\gamma_5$ matrices, we have discussed that this procedure is however not sufficient in the case of IReg. We have chosen one of the most subtle examples in the literature for illustration, the abelian chiral anomaly. Symmetrization of the trace\footnote{In the language of Appendix~\ref{sec:gamma5}, this is equivalent to using eq.\ref{eq:def G5} before evaluating the trace over the Dirac matrices.} in this case provides the correct positioning of the anomaly in the WI, in accordance with momentum routing invariance (or the vanishing of the ST) in gauge invariant processes.

\addcontentsline{toc}{section}{bibliography}
\bibliographystyle{unsrt}
\bibliography{final_v2.bib}

\end{document}